\documentclass{iopart}
\usepackage{graphicx}
\usepackage{amsfonts}
\usepackage{amssymb}
\usepackage{color}
\usepackage{psfrag}

\def\be{\begin{equation}}
\def\ee{\end{equation}}
\def\ba{\begin{eqnarray}}
\def\ea{\end{eqnarray}}

\def\modif#1{#1}

\begin{document}
\title[Diverging conductance between random and pure XX spin chains]%
      {Diverging conductance at the contact between random
        and pure quantum XX spin chains}

\author{Christophe Chatelain}
\address{
Groupe de Physique Statistique,
D\'epartement P2M,
Institut Jean Lamour (CNRS UMR 7198),
Universit\'e de Lorraine, France}

\ead{christophe.chatelain@univ-lorraine.fr}

\date{\today}

\begin{abstract}
  A model consisting in two quantum XX spin chains, one homogeneous and
  the second with random couplings drawn from a binary distribution, is
  considered. The two chains are coupled to two different non-local
  thermal baths and their dynamics is governed by a Lindblad equation. In
  the steady state, a current $J$ is induced between the two chains by
  coupling them together by their edges and imposing different chemical
  potentials $\mu$ to the two baths. While a regime of linear characteristics
  $J$ versus $\Delta\mu$ is observed in the absence of randomness, a gap
  opens as the disorder strength is increased. In the infinite-randomness
  limit, this behavior is related to the density of states of the localized
  states contributing to the current. The conductance is shown
  to diverge in this limit.
\end{abstract}


\section{Introduction}
Disorder can have drastic consequences, especially in low-dimensional
systems. At thermal equilibrium, the case of the quantum Ising chain
in a transverse field has been particularly studied.
At the quantum phase transition of the model,
the introduction of random couplings leads to a new critical behavior
governed by a new infinite-randomness fixed point~\cite{Fisher,Igloi}.
Even outside the critical point, in the so-called Griffiths phases,
rare disorder configurations induce singularities of the free energy,
as well as anomalous dynamical properties~\cite{Huse}.
\\

Much less is known about open random spin chains. In this work, we are
interested in transport properties of a random spin-$1/2$ XX chain. Usually,
a non-equilibrium steady state (NESS) is induced by coupling two different
baths to the two edges of the chain. As in the pure
case~\cite{Antal}, a non-uniform magnetization profile is expected in
the NESS. Since the XX chain can be mapped onto a free fermion gas, the
system can be seen as a tight-binding model with random couplings and
coupled to two thermal baths with different temperatures and/or
different chemical potentials. In the absence of disorder, a uniform
current is expected to flow through out the system.
Because the eigenmodes of a translation-invariant Hamiltonian are
delocalized over the whole chain, the conduction is ballistic. Any
excitation created by one bath reaches the second one. The conductivity
is therefore infinite, though the conductance is finite and
quantized~\cite{Landauer,Datta,Kawabata}. In contrast, in the presence
of randomness, the excitations are scattered on the impurities so that
the conduction becomes diffusive as the disorder is increased. 
In one-dimensional systems, the consequence is even more drastic because all
eigenstates of the Hamiltonian are localized, even with an infinitesimal
disorder. Therefore, the excitations cannot propagate from one bath to
the other. The current dies off exponentially over the so-called Anderson
localization length. The chain is an insulator~\cite{Anderson,Kramer,Evers}.
This mechanism of localization is preserved when interactions
between the charge carriers are considered, i.e. in the so-called
Many-Body Localization phase~\cite{Basko}.
\\

The coupling with a bath requires some approximations. The simplest one
is to consider two semi-infinite chains initially prepared in different
thermal states. The two chains are then coupled together and the whole
system is considered as isolated in the calculation of its time evolution.
\modif{
  In the long-time limit, the magnetization profile $m(x,t)$ was shown to be
  a function of $x/t$ for both the XX~\cite{Antal} and XXZ~\cite{Bertini}
  chains.}
The coupling of a XX chain with two baths at its edges was later
taken into account by repeatedly bringing thermalized spins in contact
with the system during short time steps~\cite{Platini}. A different
route is provided by the Lindblad equation~\cite{Breuer,Lindblad}
    \be{d\rho\over dt}=-{i\over\hbar}[H,\rho]+2D[\rho]
    \label{EqLindblad}\ee
which gives the effective time evolution of the system in the
Markovian and rotating-wave approximations. $\rho$ is the reduced density
matrix of the chain and the dissipator $D[\rho]$ has the general structure
  \be 2D[\rho]=\sum_{\alpha} \gamma_{\alpha}(t)\Big[V_{\alpha}(t)\rho(t)
  V_{\alpha}^+(t)-{1\over 2}\{V_{\alpha}^+(t)V_{\alpha}(t),\rho_{S}(t)\}\Big].\ee
When a bath is coupled only to an edge of the chain, the
two only possible Lindblad operators $V_\alpha$ are the
ladder operators $\sigma_{1}^-$, $\sigma_{1}^+$ at the left boundary and
$\sigma_{L}^-$, $\sigma_{L}^+$ at the right one. Exact results
are known for this configuration for the pure XX chain~\cite{Karevski}.
It was later shown that the steady state of the Lindblad equation admits
a representation in terms of Matrix Product State for both the
XX~\cite{Znidaric2} and the XXZ~\cite{Popkov} chains.
In molecular nanowires for which the dynamics of electrons is well
described by a tight-binding model, a third reservoir coupled to all
sites of the chain is often introduced to describe phase-breaking processes,
mainly due to electron-phonon coupling~\cite{Galperin}. Such a dephasing
effect was also studied for the XX chain by introducing a Lindblad operator
$\sigma_{i}^z$ coupled to all sites $i$ of the chain~\cite{Znidaric,Monthus}.
Interestingly, the mechanism of Anderson localization may be avoided
with such a bath because the phase decoherence is destroyed~\cite{Yusipov}.
\\

In this study, a random XX chain coupled to delocalized degrees
of freedom acting, not only as a thermal bath, but also as a reservoir
in the fermion picture, is considered. The coupling is implemented
in the Lindblad equation using the phenomenological non-local dissipator
recently introduced by Guimar\~aes {\sl et al.}~\cite{Pedro}.
In the first section, the model is discussed and the dissipator
implementing the coupling with the baths is introduced. In the second
section, the case of two homogeneous subchains is studied as a preliminary.
The conductance, not considered in \cite{Pedro}, is computed in
the limit of a weak interchain coupling. In the third section, the
case of a random left subchain is considered. A divergence of the
conductance as the disorder strength increases is observed in the
numerical simulations and explained in the infinite-randomness limit.
Conclusions follow.

\section{Description of the model}
We consider an open spin-$1/2$ XX chain governed by the Hamiltonian
\be
    H=-{1\over 2}\sum_{n=1}^{L-1} J_n(\sigma^x_n\sigma^x_{n+1}
    +\sigma^y_n\sigma^y_{n+1})-\sum_{n=1}^{L} h_n\sigma^z_n.\ee
The Jordan-Wigner transformation maps the XX chain onto a fermionic
tight-binding model, whose Hamiltonian is, up to a constant,
\be
    H=-\sum_{n=1}^{L-1} J_n(c_n^+c_{n+1}+c_{n+1}^+c_n)-2\sum_{n=1}^{L} h_nc_n^+c_n
\ee
where the exchange couplings $J_n$ play the role of hopping constants
between nearest-neighboring sites and the transverse fields $h_n$ of a
chemical potential. Such a model is relevant for the description
of conduction in polymers as polyaniline~\cite{Schulz}. In the homogeneous
case, i.e. $J_n=J$ and $h_n=h$, the unitary transformation
    \be\eta_k=\sum_{n=1}^L U_{k,n}c_n
    =\sqrt{2\over L+1}\sum_{n=1}^L\sin {nk\pi\over L+1}\ \!c_n,
    \quad (k=1\ldots L)\label{TransfoPure}\ee
diagonalizes the Hamiltonian:
    \be H=\sum_{k=1}^L \varepsilon_k \eta_k^+\eta_k\ee
with the dispersion law $\varepsilon_k=-2J\cos{\pi k\over L+1}-2h$
($k=1,\ldots,L$).    
\\

Two such XX chains are joined together to form a single chain of length
$2L$. In the right subchain, the couplings are homogeneous, $J_n=1$, while
in the left one, they are random variables drawn from the binary distribution
   \be\wp(J_n)={1\over 2}\delta\big(J_n-r\big)
   +{1\over 2}\delta\big(J_n-1/r\big).\ee
This particular choice is motivated by the fact that it allows for the
numerical computation of average quantities over all disorder
configurations so that the contribution of rare events is properly taken
into account. For a left subchain of $L$ sites, the number of random
couplings is indeed $L-1$ so that the number of disorder configurations
is $2^{L-1}$. A Jordan-Wigner transformation is performed independently in
each subchain, leading to two species of fermions $c_{\alpha,n}$ ($\alpha=1,2$).
The resulting tight-binding Hamiltonian is diagonalized and new fermionic
operators $\eta_{\alpha,k}$, annihilating a particle in the $k$-th eigenmode
of the subchain $\alpha$, are introduced. The right subchain being
homogeneous, the unitary transformation between $c_{2,n}$ and $\eta_{2,k}$ is
given by Eq.~(\ref{TransfoPure}). For the left subchain, the unitary
transformation
   \be\eta_{1,k}=\sum_{n=1}^L V_{k,n} c_{1,n}\label{Modes}\ee
is determined numerically for each disorder realization $\{J_n\}$.
The left and right subchains are then connected by a coupling $g$. The
full Hamiltonian reads
   \ba H&=&-\sum_{n=1}^{L-1} J_n (c_{1,n}^+c_{1,n+1}+c_{1,n+1}^+c_{1,n})
   -g (c_{1,L}^+c_{2,1}+c_{2,1}^+c_{1,L}) \nonumber\\
   &&\quad-\sum_{n=1}^{L-1}(c_{2,n}^+c_{2,n+1}+c_{2,n+1}^+c_{2,n}).\ea
No transverse field has been included since the chemical potential will
be fixed in the following by the two thermal baths.
Note that this model is not equivalent to Anderson model~\cite{Anderson}
for which the hopping $J_n$ is constant while the energy $h_n$ of the
fermion on each site is random.
\\

The dynamics is governed by a Lindblad equation~(\ref{EqLindblad})
with a non-local dissipator recently introduced for the homogeneous XX
chain~\cite{Pedro,Santos}. Each subchain is coupled to a different bath.
The latter are coupled to each eigenmodes $k$ of the subchains via
two Lindblad operators corresponding to the fermionic creation/annihilation
operators $\eta_{\alpha,k}^+$ et $\eta_{\alpha,k}$.
The dissipators read explicitly
  \ba D_\alpha[\rho]&=&2\gamma\sum_{k} n_{\alpha,k}
  \Big(\eta_{\alpha,k}^+\rho\eta_{\alpha,k}
  -{1\over 2}\{\eta_{\alpha,k}\eta_{\alpha,k}^+,\rho\}\Big) \nonumber\\
  &&\quad +2\gamma\sum_{k} (1-n_{\alpha,k})\Big(\eta_{\alpha,k}\rho\eta_{\alpha,k}^+
  -{1\over 2}\{\eta_{\alpha,k}^+\eta_{\alpha,k},\rho\}\Big).
  \label{Dissipator}
  \ea
The constants $n_{\alpha,k}$ are fixed by the requirement that, in the absence
of interchain coupling ($g=0$), each subchain will thermalize independently.
In the steady state, the average number of fermions in each mode is
given by the Fermi-Dirac distribution:
    \be \langle \eta_{\alpha,k}^+\eta_{\beta,k'}\rangle_{\rm st.}
    =n_{\alpha,k}\delta_{\alpha,\beta}\delta_{k,k'}
    ={\delta_{\alpha,\beta}\delta_{k,k'}\over
      e^{(\varepsilon_{\alpha,k}-\mu_\alpha)/k_BT_\alpha}-1}\ee
where $T_\alpha$ and $\mu_\alpha$ are respectively the temperature and the
chemical potential of the bath and $\varepsilon_{\alpha,k}$ is the energy
of the $k$-th eigenmode of the isolated subchain.
The dynamical equation of the equal-time correlation function
$C_{\alpha,k;\beta,k'}=\langle \eta_{\alpha,k}^+(t)\eta_{\beta,k'}(t)\rangle$ can
be cast into the convenient matrix form~\cite{Pedro}
    \be {d\over dt}C={i\over\hbar}[W_0+W_1,C]-\{\Gamma,C\}+2{\cal D}
    \label{EqEvolCorr}\ee
where $W_0$, $\Gamma$ and ${\cal D}$ are diagonal matrices whose entries
are $\varepsilon_{\alpha,k}$, $\gamma$ and $\gamma n_{\alpha,k}$ respectively.
The non-diagonal matrix $W_1$ comes from the interchain coupling.
In the following, Eq.~(\ref{EqEvolCorr}) is integrated numerically using
a 4th-order Runge-Kutta algorithm up to a time sufficiently large ($t=100$)
for the steady state to be reached.
\\

The particle current between the two subchains reads~\cite{Pedro}
   \ba J&=&-ig\big(\langle c_{1,L}^+c_{2,1}\rangle
   -\langle c_{2,1}^+c_{1,L}\rangle\big) \nonumber\\
   &=&2g\Im \sum_{k,k'} V_{L,k}^*U_{1,k'}
   \langle\eta^+_{1,k}\eta_{2,k'}\rangle.\label{Current}\ea
In the steady state this current is equal to the current between
the first subchain and the bath with which it is coupled:
  \be J=-2\gamma\sum_k \big(n_{1,k}-\langle c_{1,k}^+c_{1,k}\rangle\big).\ee
The particle current between the two subchains is measured numerically and
averaged over all disorder configurations. The model will also be studied
analytically in the limit of a weak interchain coupling $g$. To lowest
order in $g$, the current~(\ref{Current}) reads~\cite{Pedro}
   \be J=-4\gamma g^2\sum_{k,k'}|V_{k;L}|^2|U_{k';1}|^2
       {\big(n_{2,k'}-n_{1,k}\big)\over 4\gamma^2
         +\big(\varepsilon_{1,k}-\varepsilon_{2,k'}\big)^2}+{\cal O}(g^3)
       \label{Current2}\ee
The limit of an infinitely strong disorder, i.e. $r\rightarrow +\infty$,
will be discussed.

\section{Conductance between two pure subchains}
We first consider two pure chains with $J_n=1$. In the usual setup where
the chain is coupled to two baths at its edges, the conductivity is infinite
because the propagation of excitations (fermions) is ballistic. Any fermion
of wavevector $k>0$ injected in the system at the left boundary reaches
the right boundary. In the system considered in this work, fermions introduced
by the left bath into the left subchain can be scattered by the coupling
between the two subchains:
  \be\fl -g (c_{1,L}^+c_{2,1}+c_{2,1}^+c_{1,L})
  =-g\sum_{k,k'} \big(U_{kL}U_{k'1}^*\eta_{1,k}^+\eta_{2,k'}
  +U_{kL}^*U_{k'1}\eta_{2,k'}^+\eta_{1,k}\big)\ee
and, as a consequence, reach the right subchain with a different wavevector.
In a second step (second-order in $g$), they can be scattered back to the
left subchain. The only constrain comes from the exclusion principle which
forbids the scattering of an electron from the mode $k$ in the left subchain
to the mode $k'$ in the right one if the latter is not unoccupied. Therefore,
as in the B\"uttiker formalism, the total current~(\ref{Current}) depends on
the net number of allowed scattering channels from an occupied mode in one
subchain to an unoccupied mode in the other subchain. Different chemical
potentials in left and right reservoirs induce an imbalance and, as a
consequence, a current.
\\

\begin{figure}
  \begin{center}
    \psfrag{j/g}[Bc][Bc][1][1]{$J/g^2$}
    \psfrag{2}[Bc][Bc][1][1]{}
    \psfrag{dmu}[tc][tc][1][0]{$\Delta\mu$}
    \psfrag{  T=0.001}[Bc][Bc][1][1]{\tiny $T=0.001$}
    \psfrag{  T=0.02}[Bc][Bc][1][1]{\tiny $T=0.02$}   
    \includegraphics[width=6.37cm]{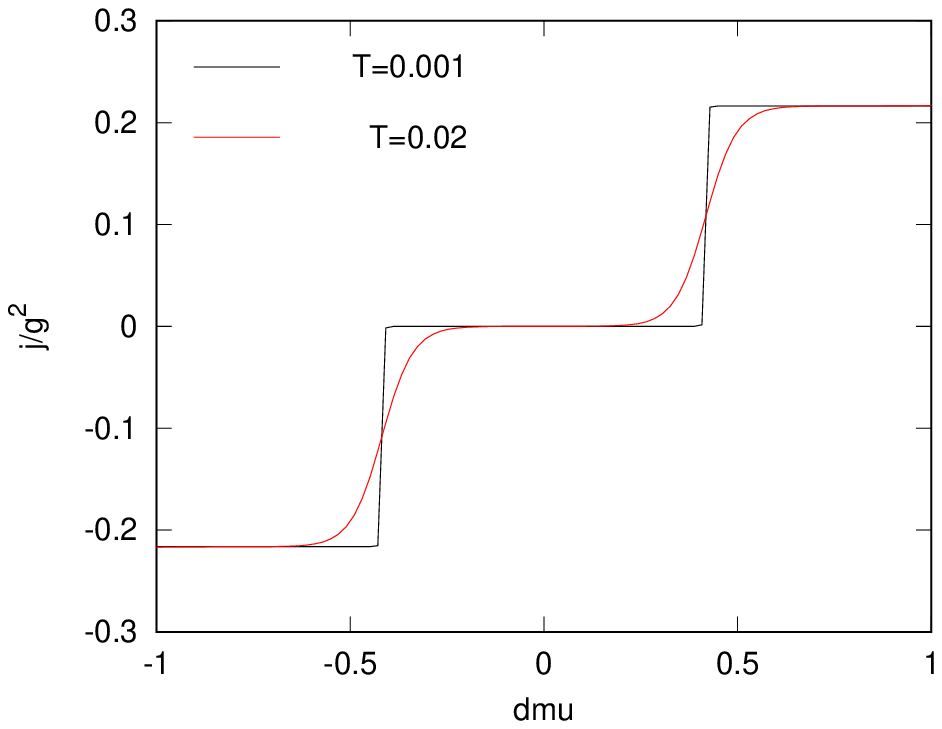}
    \includegraphics[width=6.37cm]{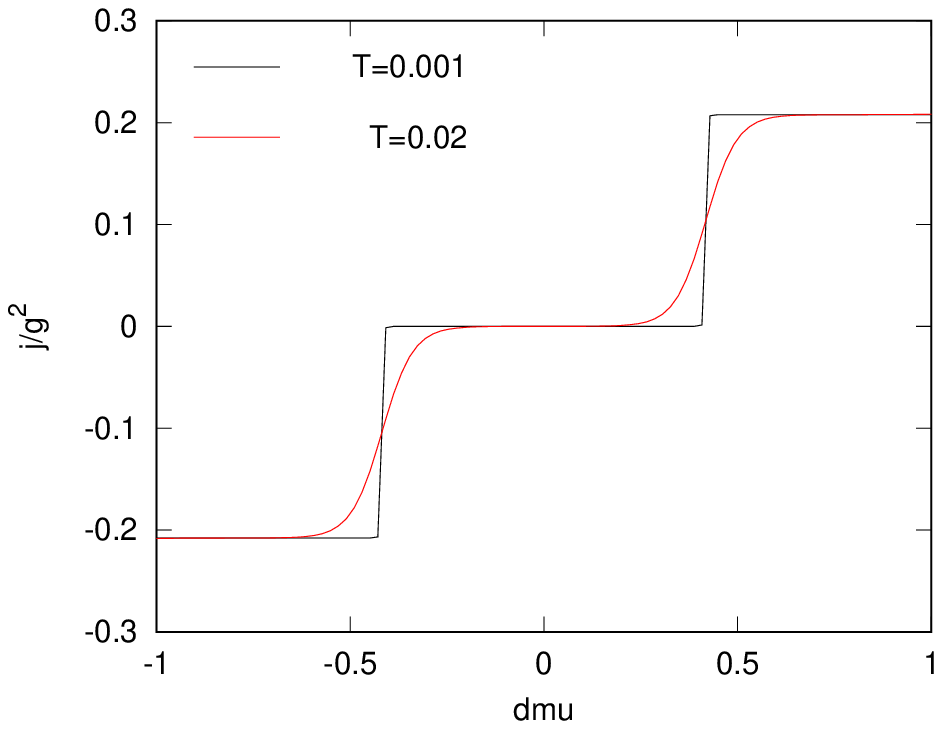}\\
    \includegraphics[width=6.37cm]{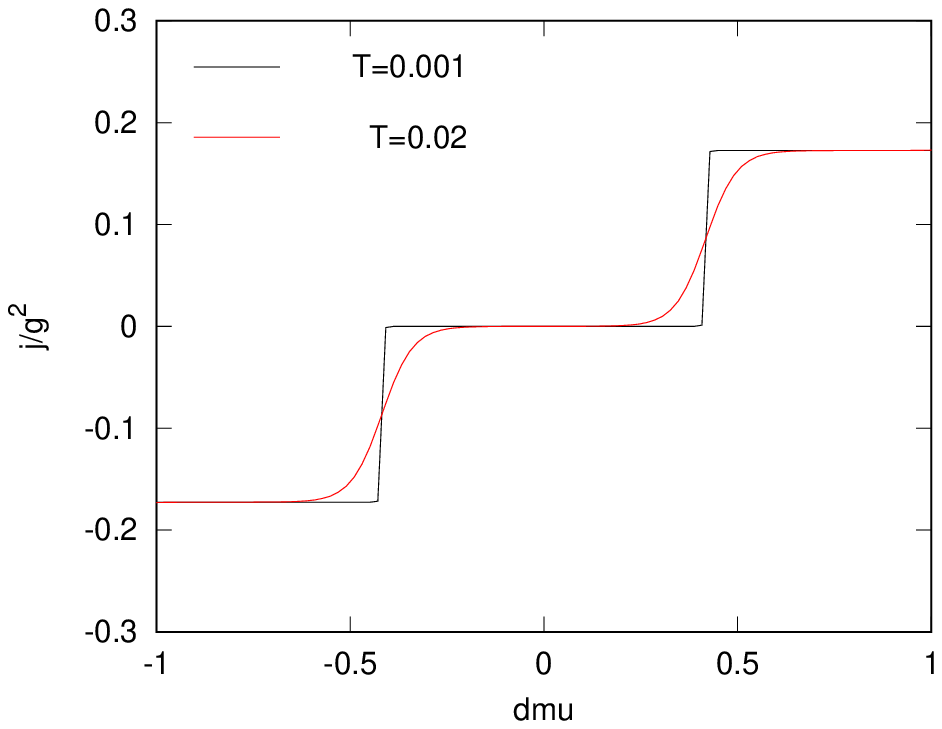}
    \caption{Current $J/g^2$ between two pure XX chains of size $L=14$
      versus $\Delta\mu$ when the two chains are in contact with baths at
      a different chemical potential $\Delta\mu/2$ and $-\Delta\mu/2$.
      The two curves corresponds to different temperatures $T=0.001$ and
      $T=0.02$. The interchain coupling is $g=0.1$ (top left),
      $0.3$ (top right), and $0.7$ (bottom).}
  \label{fig1}
  \end{center}
\end{figure}

\begin{figure}
  \begin{center}
    \psfrag{j}[Bc][Bc][1][1]{$J/g^2$}
    \psfrag{dmu}[tc][tc][1][0]{$\Delta\mu$}
    \psfrag{L=14}[Bc][Bc][1][1]{\tiny $L=14$}
    \psfrag{L=28}[Bc][Bc][1][1]{\tiny $L=28$}
    \psfrag{L=56}[Bc][Bc][1][1]{\tiny $L=56$}
    \psfrag{Inf}[Bc][Bc][1][1]{\tiny $L\rightarrow +\infty$}
   \includegraphics[width=6.37cm]{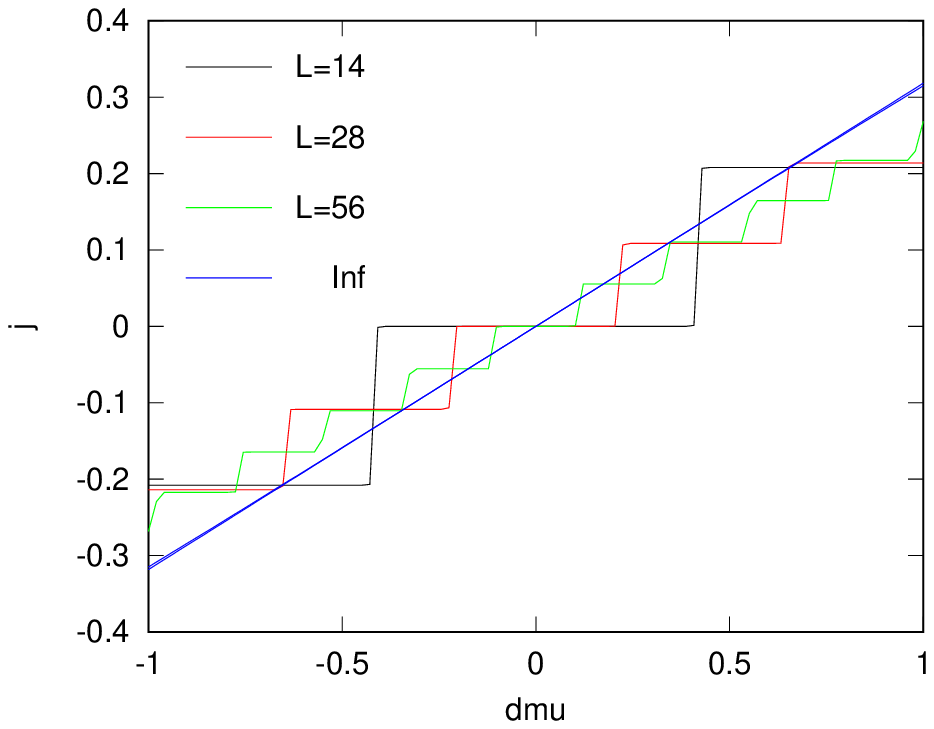}
    \includegraphics[width=6.37cm]{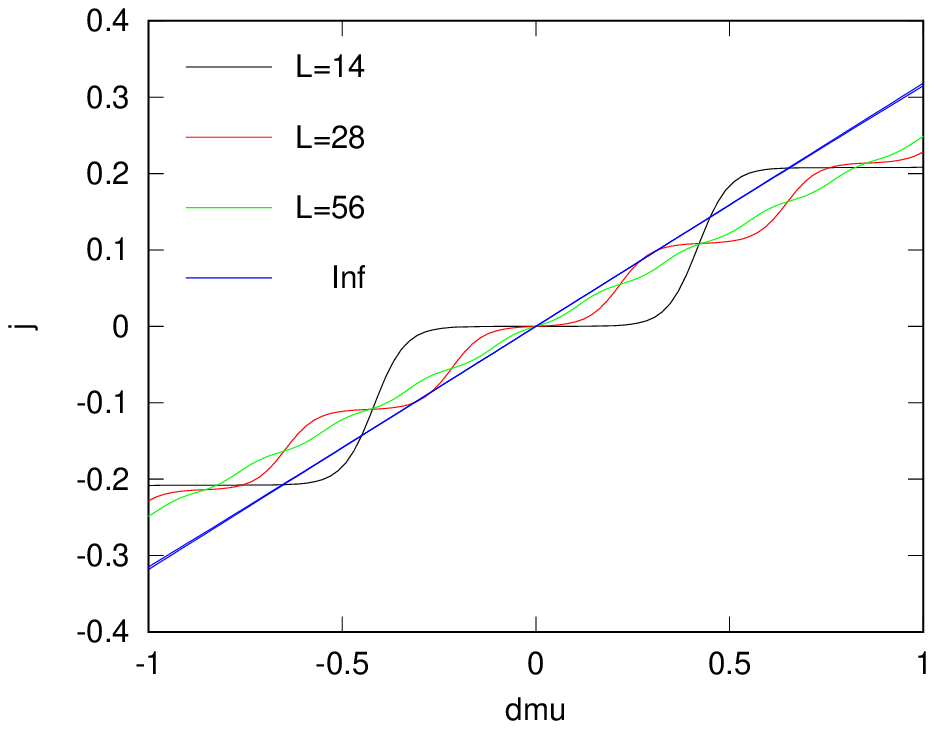}
    \caption{Current $J/g^2$ between two pure XX chains versus $\Delta\mu$
      when the two chains are in contact with baths at a different chemical
      potential $\Delta\mu/2$ and $-\Delta\mu/2$. The different curves
      correspond to different lattice sizes $L$ and, the last one, to
      the analytical expression (\ref{CurrentInfRPure}).
      On the left graph, the temperature of the left and right baths is
      $T=0.001$ while it is equal to $0.02$ on the right.}
  \label{fig2}
  \end{center}
\end{figure}

Numerically, the current is observed to display a staircase-like behavior
with the difference of chemical potentials $\Delta\mu$ between the two
reservoirs at low temperature (figure~\ref{fig1}). These steps are due to
the quantization of the eigenmodes in the two finite-size subchains. They
should not be confused with the steps observed in
experiments~\cite{Wees,Wharam} where the wire is not purely
one-dimensional so that the wavevectors have a transverse component which
remains quantized when the length of the wire goes to infinity but not his
width. In the model considered here, the steps are indeed observed to vanish
as the length is increased (figure~\ref{fig2}). As expected, the steps are
smoothed as the temperature is increased. One can also notice by
comparing the left and right figures~\ref{fig1} that $J/g^2$ displays only
a small residual dependance on $g$ when $g\le 0.3$, in agreement with the
perturbative expansion~(\ref{Current2}).

The conductance between two pure subchains can be computed analytically
in some limits. As mentioned above, the coupling $g$ is treated
perturbatively. To lowest order in $g$, the current is given by
(\ref{Current2}) where $|V_{k;L}|^2=|U_{k;1}|^2={2\over L+1}
\sin^2{\pi k\over L+1}$, and $\varepsilon_{1,k}=\varepsilon_{2,k}
=-2\cos{\pi k\over L+1}$. Then, the temperature is set to zero for
both left and right reservoirs and the dissipation $\gamma$ is assumed
to be strong. In these limits, the current reads
   \be\fl J\simeq -{4 g^2\over (L+1)^2\gamma}\sum_{k,k'=1}^L
   \sin^2 {\pi k\over L+1}\sin^2 {\pi k'\over L+1}
   \big(\theta(\mu_2-\varepsilon_{2,k'})
   -\theta(\mu_1-\varepsilon_{1,k})\big).\ee
Introducing the density of states on the first (or last) site of the chain as
   \ba\rho(\varepsilon)&=&{2\over L+1}\sum_k\delta(\varepsilon-\varepsilon_{k})
   \sin^2 {\pi k\over L+1}\nonumber\\
   &\simeq& {2\over\pi}\int \delta(\varepsilon+2\cos k)\sin^2k dk\nonumber\\
   &=&{1\over\pi}\sqrt{1-\varepsilon^2/4}\ea
in the thermodynamic limit $L\rightarrow +\infty$, the current becomes
    \be J\simeq -{g^2\over\gamma}\int_{-2}^2 \rho(\varepsilon)\rho(\varepsilon')
    \big(\theta(\mu_2-\varepsilon')-\theta(\mu_1-\varepsilon)\big)
    d\varepsilon d\varepsilon'\ee
Swapping $\varepsilon$ and $\varepsilon'$ in the last term
    \be J\simeq -{g^2\over\gamma}\int_{\mu_1}^{\mu_2}\rho(\varepsilon')
    d\varepsilon'\int_{-2}^2 \rho(\varepsilon)d\varepsilon,\ee
the second integral is equal to 1. In the case $\mu_1=\Delta\mu/2$ and
$\mu_2=-\Delta\mu/2$, the current finally reads
    \be\fl J\simeq {g^2\over\pi\gamma}\int_{-\Delta\mu/2}^{\Delta\mu/2}
    \sqrt{1-\varepsilon^2/4}d\varepsilon={g^2\over\pi\gamma}\left({\Delta\mu
      \sqrt{16-\Delta\mu^2}\over 8}+2\arcsin{\Delta\mu\over 4}\right)
    \label{CurrentInfRPure}\ee
In the neighborhood of $|\Delta\mu|=0$, the current behaves as
$J\simeq {g^2\over\pi\gamma}\Delta\mu$, i.e. the conductance is
${g^2\over\pi\gamma}$. Despite the various assumptions made, the
behavior Eq.~(\ref{CurrentInfRPure}) is consistent with the numerical data
(figure~\ref{fig2}), even at non-zero temperature and for relatively small
lattice sizes.

\section{Conductance between a pure and a random system}
\subsection{Conductance at zero chemical potential}
\label{sec4.1}
\begin{figure}
  \begin{center}
    \psfrag{j/g}[Bc][Bc][1][1]{$\bar J/g^2$}
    \psfrag{2}[Bc][Bc][1][1]{}
    \psfrag{dmu}[tc][tc][1][0]{$\Delta\mu$}
    \psfrag{  T=0.001}[Bc][Bc][1][1]{\tiny $T=0.001$}
    \psfrag{  T=0.02}[Bc][Bc][1][1]{\tiny $T=0.02$}   
    \includegraphics[width=6.37cm]{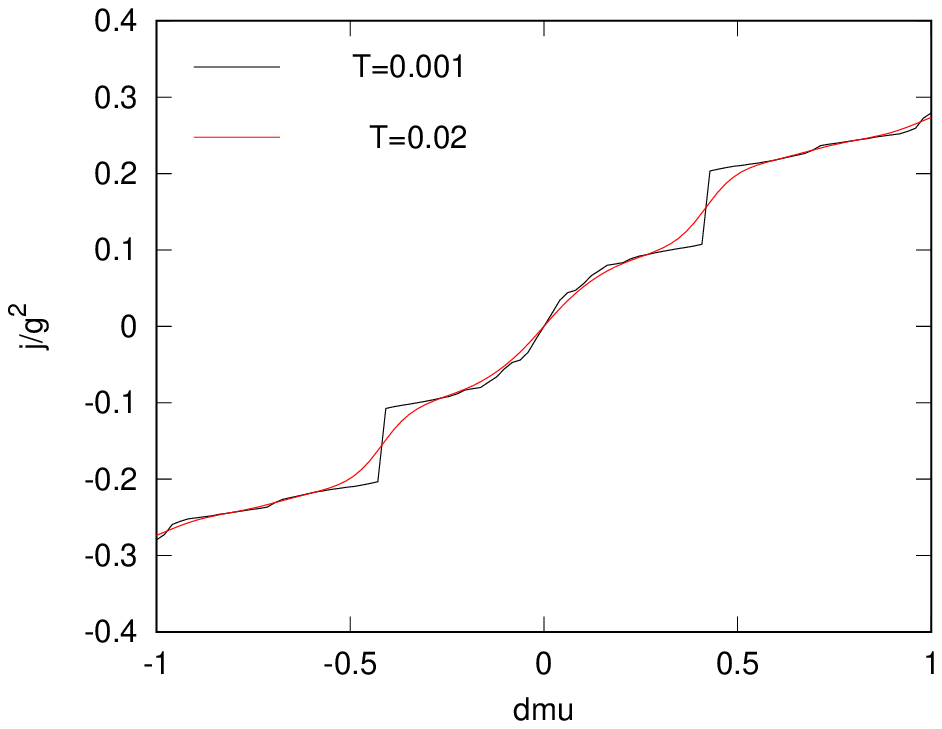}
    \includegraphics[width=6.37cm]{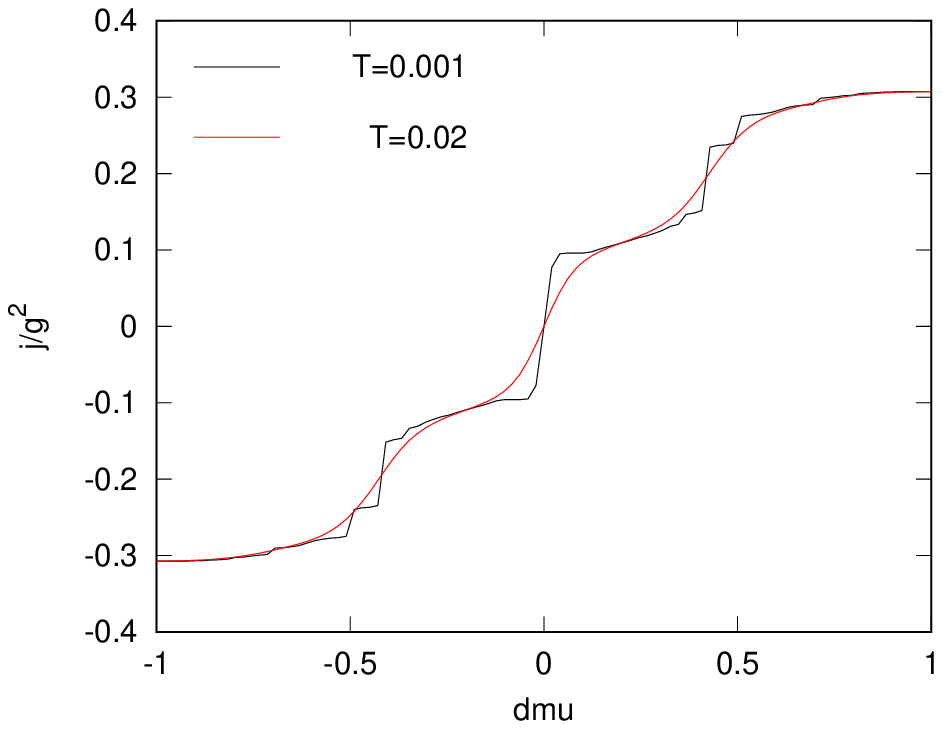}\\
    \includegraphics[width=6.37cm]{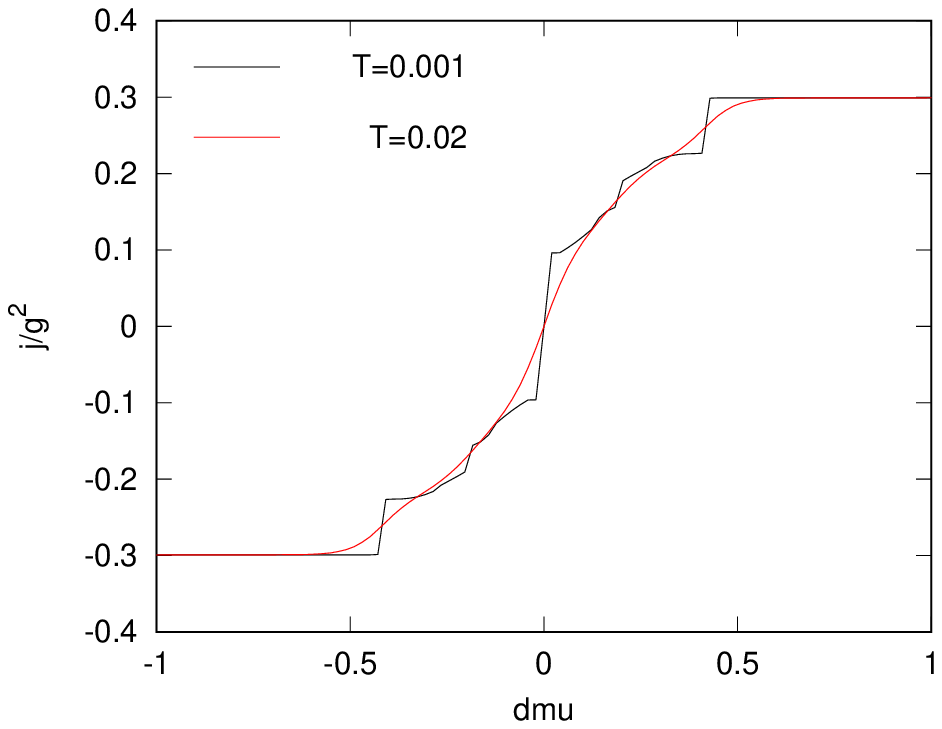}
    \caption{Average current $\bar J/g^2$ between a random and a pure
      XX chains of size $L=14$ versus $\Delta\mu$ when the two chains
      are in contact with baths at different chemical potentials
      $\Delta\mu/2$ and $-\Delta\mu/2$. The couplings of the random XX
      model are randomly chosen in $\{1/r;r\}$
      with $r=2$ (top left), $r=4$ (top right) and $r=10$ (bottom).
      The two curves corresponds to different temperatures $T=0.001$
      and $T=0.02$.}
  \label{fig3}
  \end{center}
\end{figure}

The right chain is still homogeneous ($J=1$) but the left one is now
random with $J_n\in \{r,1/r\}$. The temperatures of the two baths are equal
but the chemical potentials are assumed to be $\Delta\mu/2$ for the left
bath and $-\Delta\mu/2$ for the right one. The average current $\bar J$
is plotted on figure~\ref{fig3}. As in the pure case considered above,
no current flows between the two subchains if $\Delta\mu=0$. For the
weakest disorder ($r=2$), the curve displays a shape still roughly similar
to the pure case (figure~\ref{fig1}). However, as the strength of disorder
increases, the curve becomes steeper and steeper at $\Delta\mu=0$.
The conductance, defined as the derivative~\cite{Landauer}
      \be G=\left({d\bar J\over d\Delta\mu}\right)_{\Delta\mu=0},\ee
increases with $r$. From the curve corresponding to the strongest disorder
($r=10$), one can conjecture that a gap will open at zero temperature in
the limit of infinite randomness, yielding in this case an infinite
conductance. Both finite disorder and finite temperature smooth the curve
and make the conductance finite.

\begin{figure}
  \begin{center}
    \psfrag{Pr(E)}[Bc][Bc][1][1]{$\rho(\varepsilon)$}
    \psfrag{E}[Bc][Bc][1][1]{$\varepsilon$}
    \includegraphics[width=6.37cm]{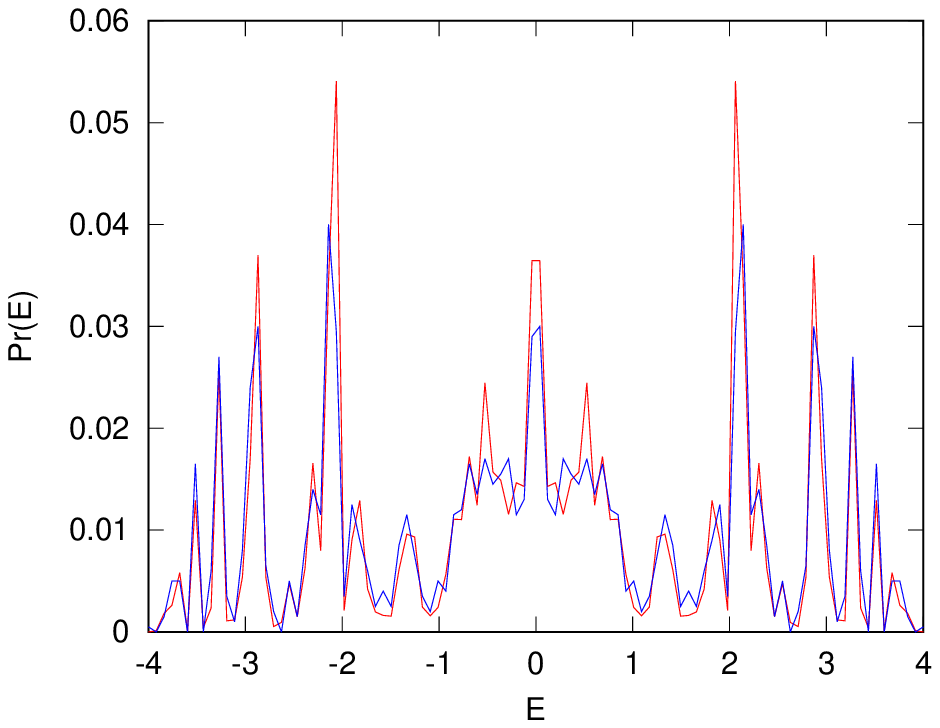}
    \includegraphics[width=6.37cm]{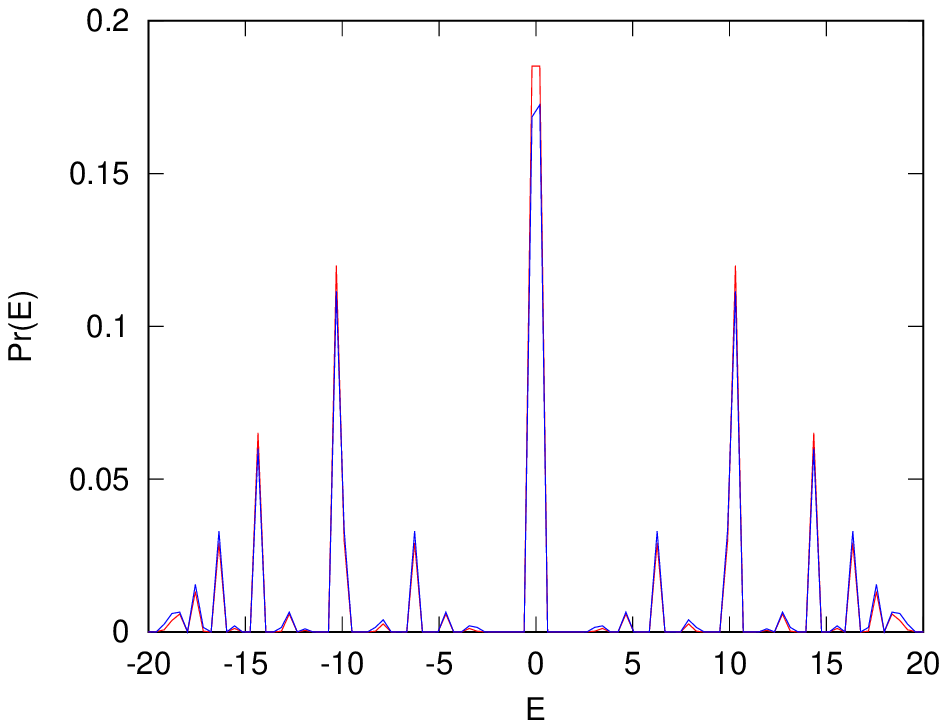}
    \caption{Probability density $\wp(\varepsilon)$ of the eigenvalues of the
      random tight-binding Hamiltonian with $r=2$ (left) and $r=10$ (right).
      The red curve corresponds to the distribution for a lattice size $L=16$
      averaged over the $32768$ random configurations of the couplings.
      The blue curve is the probability density for a single random realization
      of a large lattice of size $L=2000$.}
  \label{fig4}
  \end{center}
\end{figure}

The current being related to the imbalance between the number of scattering
channels from left to right and right to left, an explanation for the opening
of a gap observed numerically should be looked for in the energy spectrum of
the random XX subchain. As often noticed in the literature, the level spacing
distribution of the random XX subchain is not compatible with the Gaussian
Orthogonal Ensemble observed in chaotic systems~\cite{Bohigas, Beenakker}.
It can be checked numerically
that the levels are distributed according to a Poisson distribution
$\wp(\Delta E)\sim e^{-\Delta E}$ in the random XX chain. However, this
distribution is not relevant to our purpose because the current is expressed
in terms of fermionic excitations. More relevant here is the density of
states of the free fermion gas after the Jordan-Wigner transform of the XX
model. The latter is presented on figure~\ref{fig4}. For strong randomness,
the density of states appears as a sequence of Dirac peaks, the largest
one being located at $\varepsilon=0$. This behavior is easily understood
in the  infinite-randomness limit $r\rightarrow +\infty$. For a given disorder
realization, the fermion Hamiltonian is a tridiagonal matrix whose elements
are $H_{n,n+1}=H_{n+1,n}=J_n$ where $J_n$ is a random variable in $\{r,1/r\}$.
In the infinite-randomness limit $r\rightarrow +\infty$, at least one fourth of
the eigenvalues of the Hamiltonian are expected to vanish. Indeed, the
probability that two consecutive couplings, say $J_n$ and $J_{n+1}$, are
both equal to $1/r$, and therefore vanish in the limit $r\rightarrow +\infty$,
is $1/4$. As a consequence, the $(n+1)$-th line of the matrix $H-\varepsilon$
has only one non-zero element, $-\varepsilon$ on the diagonal. The determinant
of the $L\times L$ matrix $H-\varepsilon$ therefore reads $(-\varepsilon)
M_{n+1,n+1}$ where $M_{n+1,n+1}$ is the minor of the matrix. The characteristic
polynomial of the matrix can be factorized by $-\varepsilon$ which implies
that $\varepsilon=0$ is one of its roots.
When $N$ consecutive couplings are large and bounded by two weak couplings,
i.e. $J_n=J_{n+N+1}=1/r$ and $J_{n+m}=r$ for any $m=1,\ldots, N$, the $N+1$
sites $n+1$ to $n+N+1$ are uncoupled from the rest of the system in the
infinite-randomness limit. For a given disorder realization, the fermion
Hamiltonian is a block diagonal matrix. Each block of $N$ large couplings
corresponds to the Hamiltonian of a fermion on a chain of $N+1$ sites with
a hopping constant $r$. The $N+1$ eigenvalues are~\cite{Nechaev}
  \be\varepsilon_{N,q}=-2r\cos{\pi q\over N+2},\quad
  q=1,\ldots, N+1\label{BlockEig}\ee
and the associated eigenvectors
  \be\phi_{N,q}(m)=\sqrt{2\over N+2}\sin{\pi qm\over N+2},\quad
  m=1,\ldots, N+1.\label{BlockEVec}\ee
The eigenvectors have zero component on other sites. The probability to
observe such a block is $1/2^{N+2}$, i.e. the localization length is
$\xi=1/\ln 2$. The density of states is not a smooth function, even in
the thermodynamic limit, but displays a discrete number of peaks.
Each one of these peaks can be associated to two numbers, $N\in\mathbb{N}$
and $q=1,\ldots, N+1$. This picture is consistent with what is observed
numerically on figure \ref{fig4}.
\\

Coming back to the expression (\ref{Current2}) of the current, one can notice
that the scattering of an eigenmode $k$ of the left subchain to an eigenmode
$k'$ of the right one is weighted by $|V_{k;L}|^2|U_{k';1}|^2$, i.e. the
probability to find the fermion in the eigenmode $k$ on site $L$ of the left
subchain times the probability to find the fermion in the eigenmode $k'$
on site $1$ of the right one. Since the eigenmodes of the pure right
subchain are delocalized, $|U_{k';1}|^2={2\over L+1}\sin^2 {\pi k'\over L+1}$
does not vanish. In contrast, in
the random left subchain, all states are localized in the thermodynamic
limit. Therefore, only those states with a non-zero probability on the
site $L$ will contribute to the current. These states are surface states
and were not discussed above. If the last coupling $J_{L-1}=1/r$ is weak
then, in the infinite-randomness limit, the site $L$ is decoupled from the
rest of the chain. All elements $H_{L,n}$ of the Hamiltonian vanish. Therefore,
there exists an eigenvalue $\varepsilon=0$ associated to the eigenvector
$\phi(n)=\delta_{n,L}$. The probability of such an event is $1/2$. In a way
analogous to what happens in the bulk, when the $N$ last couplings are equal
to $r$, i.e. $J_{L-N}=\ldots=J_{L-1}=r$ and $J_{L-N-1}=1/r$, then,
in the infinite-randomness limit, the last $N+1$ sites are decoupled from
the rest of the system. A $(N+1)\times (N+1)$ block appears on the diagonal
of the Hamiltonian. After diagonalization, the $N+1$ eigenvectors are found
to be given by (\ref{BlockEVec}) with the associated energies
(\ref{BlockEig}). The probability of such a block is $1/2^{N+1}$.

At zero-temperature, the average current reads
  \be\fl\bar J=-4\gamma g^2\sum_{k,k'}\overline{|V_{k;L}|^2|U_{k';1}|^2
  {\theta(\mu_2-\varepsilon_{2,k'})-\theta(\mu_1-\varepsilon_{1,k})
    \over 4\gamma^2+\big(\varepsilon_{1,k}-\varepsilon_{2,k'}\big)^2}}
  +{\cal O}(g^3).  \label{CurrentLowT}\ee
All energies $\varepsilon_{1,k}$ (\ref{BlockEig}) are proportional to $r$,
and therefore diverge in the infinite-randomness limit, apart from the
surface states for which $\varepsilon=0$. As a consequence, only those
latter states contribute to the current (\ref{CurrentLowT}). The energies
$\varepsilon_{N,q}$ vanish when the fermion is localized on the site $L$
of the left chain or in a block of $N$ strong couplings when $N$ is even
and $q={N\over 2}+1$. The square of the wavefunction on the right site,
i.e. $|V_{k;L}|^2$, is equal to one in the former case and to
   \be|\phi_{N,q={N\over 2}+1}(N)|^2={2\over N+2}
   \sin^2\left({\pi Nq\over N+2}\right)
    ={2\over N+2}\ee
in the latter. The probability of such a block being $1/2^{N+1}$,
the weight appearing in the current is
   \be{1\over 2}+\sum_{N=2\atop\rm even}^{+\infty}
   {|\phi_{N,q={N\over 2}+1}(N)|^2\over 2^{N+1}}
   ={1\over 2}+\sum_{N=2\atop\rm even}^{+\infty}{1\over (N+2)2^{N}}
   =2\ln{4\over 3}.\ee
The average current is now
   \be\fl\bar J=-8\gamma g^2\ln{4\over 3}
   \sum_{k'} {2\over L+1}\sin^2\left({\pi k'\over L+1}\right)
   {\theta(\mu_2-\varepsilon_{2,k'})-\theta(\mu_1)
     \over 4\gamma^2+(\varepsilon_{2,k'})^2}+{\cal O}(g^3).\ee
The sum is evaluated in the thermodynamic limit $L\rightarrow +\infty$.
When $\mu_1<0$, the left states $\varepsilon_1=0$ are empty so the current
flows from the right to the left:
    \ba\bar J(\mu_2)&=&-{16\gamma g^2\over\pi}\ln{4\over 3}
    \int_0^{k_F}{\sin^2k\over 4\gamma^2+4\cos^2k}dk
    \quad (\mu_1<0)\nonumber\\
    &=&-{4\gamma g^2\over\pi}\ln{4\over 3}
    \left[{\sqrt {1+\gamma^{2}}\over\gamma}
      \arctan{\gamma\tan k_F\over \sqrt {1+\gamma^{2}}}-k_F\right]\ \!
    \label{CurrentAn1a}
    \ea
where the Fermi wavevector satisfies $-2\cos k_F=\mu_2$ and the
$\arctan(x)$ function should be defined on $[0;\pi[$ instead of $]-\pi/2;
\pi/2[$. When $\mu_1>0$, the left states are occupied and the current
flows only from left to right:
    \ba\fl\bar J(\mu_2)&=&{4\gamma g^2\over\pi}\ln{4\over 3}
    \left[{\sqrt {1+\gamma^{2}}\over\gamma}\arctan{\gamma\tan k
        \over \sqrt {1+\gamma^{2}}}-k\right]_{k_F}^\pi\quad
    (\mu_1>0)\label{CurrentAn1b}    \ea

\begin{figure}
  \begin{center}
    \psfrag{j/g}[Bc][Bc][1][1]{$\bar J/g^2$}
    \psfrag{2}[Bc][Bc][1][1]{}
    \psfrag{dmu}[tc][tc][1][0]{$\Delta\mu$}
    \psfrag{  T=0.001}[Bc][Bc][1][1]{\tiny $T=0.001$}
    \psfrag{  T=0.02}[Bc][Bc][1][1]{\tiny $T=0.02$}   
    \psfrag{Inf}[Bc][Bc][1][1]{\tiny $r\rightarrow +\infty$}   
    \includegraphics[width=6.37cm]{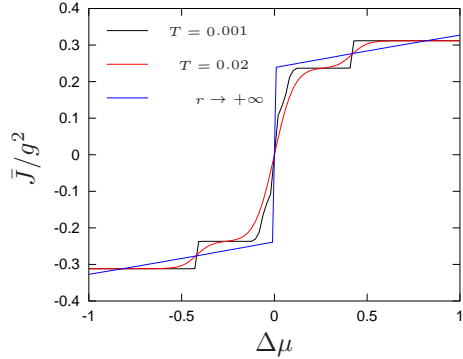}
    \caption{Average current $\bar J/g^2$ between a random and a pure XX chain
      of sizes $L=14$ versus $\Delta\mu$ when the two chains are in contact with
      baths at a different chemical potentials $\Delta\mu/2$ and $-\Delta\mu/2$.
      The couplings of the random XX model are randomly chosen in $\{1/r;r\}$
      with $r=30$ and the inter-chain coupling is $g=0.1$. The two first curves
      correspond to different temperatures $T=0.001$ (black) and
      $T=0.02$ (red). The last curve (blue) is the analytical expression
      (\ref{CurrentAn1a}) and (\ref{CurrentAn1b}).
    }
  \label{fig5}
  \end{center}
\end{figure}

Considering as before $\mu_1=\Delta\mu/2$ and $\mu_2=-\Delta\mu/2$,
the current turns out to be an odd function of $\Delta\mu$ and to display
a quasi-linear dependence on $\Delta\mu$ when $\Delta\mu\sim{\cal O}(1)$ but
with a gap at $\Delta\mu=0$ equal to
    \be \Delta\bar J=4\gamma g^2\ln{4\over 3}
    \left({\sqrt {1+\gamma^{2}}\over\gamma}-1\right)\ee
Since the derivative of (\ref{CurrentAn1a}) is
    \be {d\bar J\over dk_F}={4\gamma g^2\over\pi}\ln{4\over 3}
    {\tan^2k_F\over 1+\gamma^2+\gamma^2\tan^2k_F}\label{Derivee}\ee
the slope of the current at small non-zero chemical potential is given by
    \be \lim_{k_F\rightarrow{\pi/2}^+} {d\bar J\over dk_F}{dk_F\over d\Delta\mu}
    ={4\gamma g^2\over\pi}\ln{4\over 3}\times {1\over\gamma^2}
    \times {1\over 4}\ee
when $\Delta\mu>0$. Repeating the calculation for $\Delta\mu<0$, it follows
that the average current at first order in $\Delta\mu$ is
   \be\fl\bar J={g^2\over\gamma\pi}\ln{4\over 3}\Delta\mu
   +2\gamma g^2\ln{4\over 3}
   \left({\sqrt {1+\gamma^{2}}\over\gamma}-1\right){\rm\ sign\ }\Delta\mu
   +{\cal O}(\Delta\mu^2).\label{CurrentAna3}\ee
When plotted on figure~\ref{fig5}, this expression cannot be distinguished
from equations (\ref{CurrentAn1a}) and (\ref{CurrentAn1b}). The first term
is similar to the linear behavior observed in the pure case. The slope
$\bar G={g^2\over\gamma\pi}\ln{4\over 3}$ only differs from the conductance
of the pure chain by a factor $\ln{4\over 3}$. However, the second term
of (\ref{CurrentAna3}) leads to an infinite conductance at zero chemical
potential in the infinite-randomness limit.

\subsection{Conductance at non-zero chemical potential}

In the pure case, the current $J$ between the two subchains vanishes when
the chemical potentials of the two baths are equal. This is no longer the
case when one of the chains is random. Instead, a non-zero current is
observed whenever the chemical potential in the random subchain is different
from zero.

\begin{figure}
  \begin{center}
    \psfrag{J/g^2}[Bc][Bc][1][1]{$\bar J/g^2$}
    \psfrag{2}[Bc][Bc][1][1]{}
    \psfrag{mu1}[tc][tc][1][0]{$\mu_2$}
    \psfrag{mu2}[tc][tc][1][0]{$\mu_1$}
    \includegraphics[width=6.37cm]{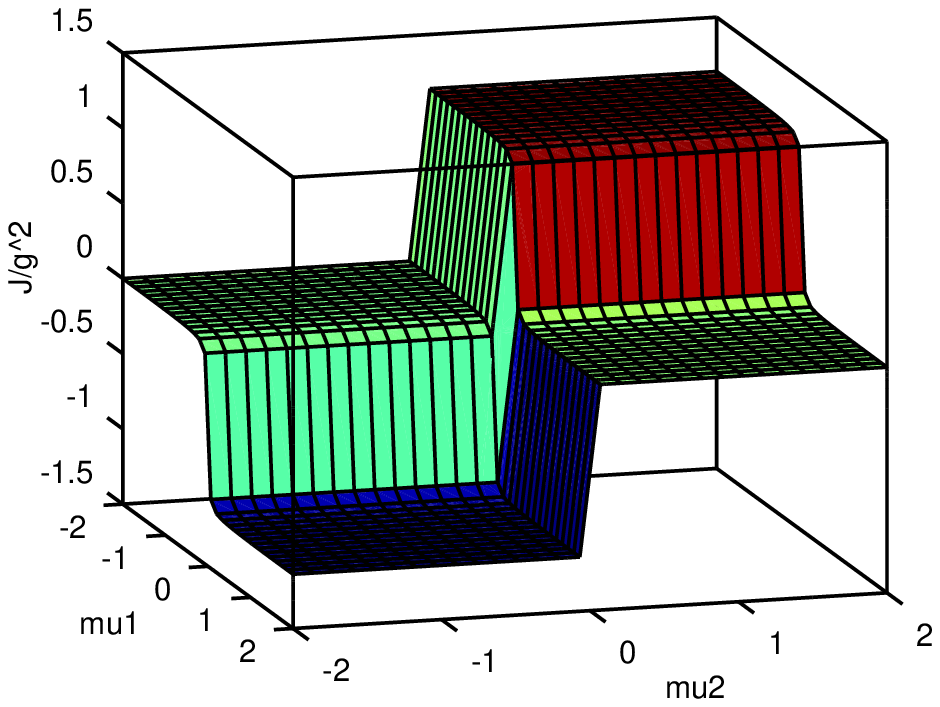}
    \includegraphics[width=6.37cm]{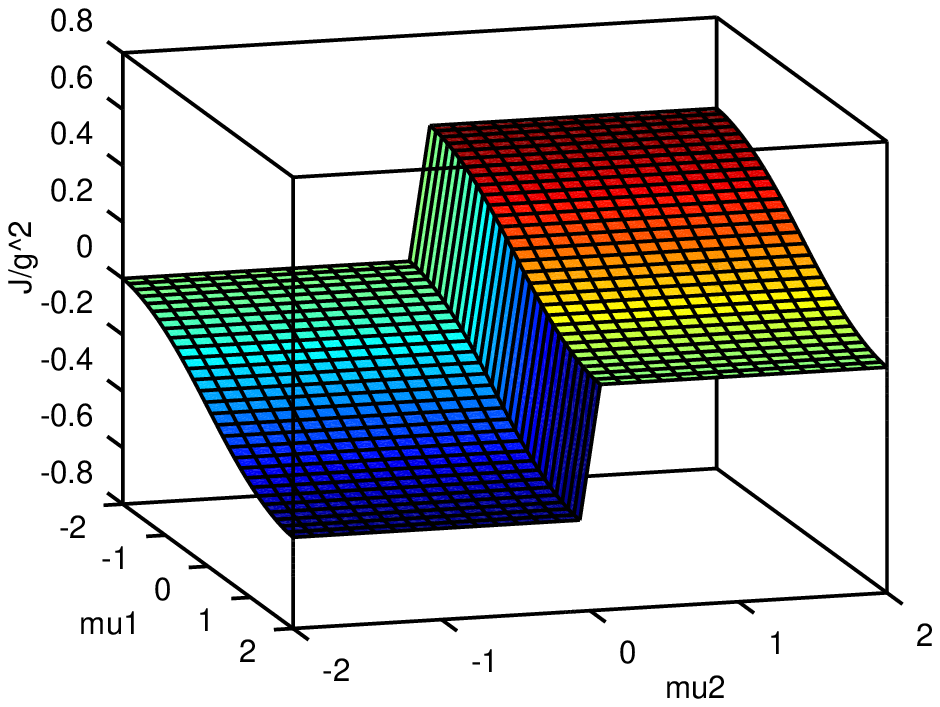}\\
    \includegraphics[width=6.37cm]{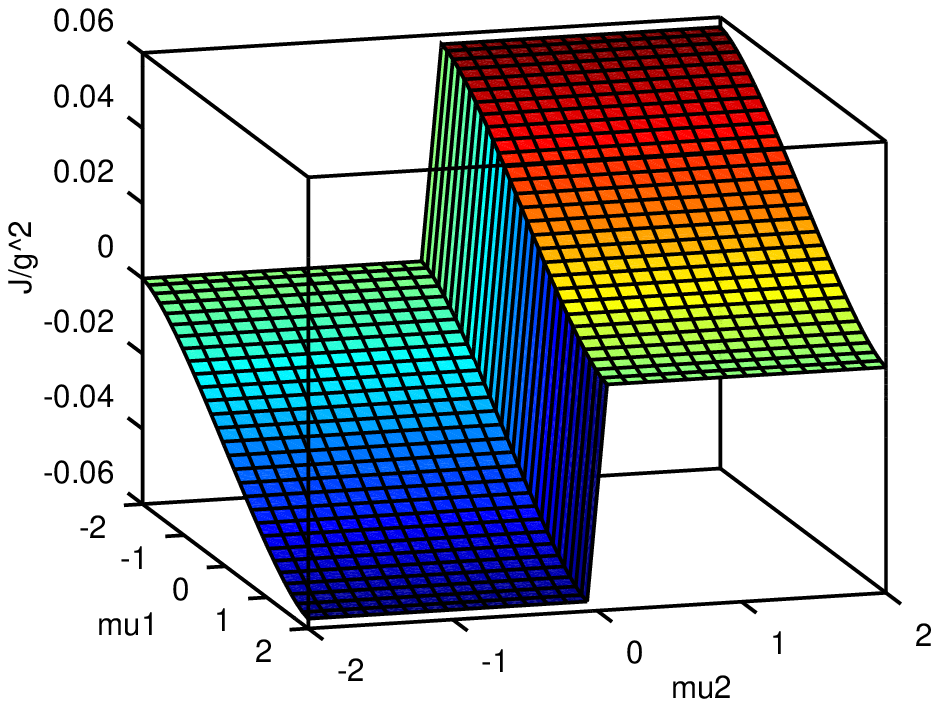}
    \caption{Average current $\bar J/g^2$ versus the chemical potentials of
      the two reservoirs in the infinite-randomness limit. The graphs
      correspond to different values of the dissipation of the baths:
      $\gamma=0.01$ (top left), $\gamma=1$ (top right),
      and $\gamma=10$ (bottom).}
  \label{fig6}
  \end{center}
\end{figure}

On figures~\ref{fig6}, the average current $\bar J(\mu_1,\mu_2)$,
as given by equations (\ref{CurrentAn1a}) and (\ref{CurrentAn1b}) in the
infinite-randomness limit, is plotted versus the chemical potentials $\mu_1$
and $\mu_2$ of the two baths in contact with the two subchains. The current
displays a monotonous dependance on $\mu_2$ but depends only on the sign of
$\mu_1$. As discussed in the previous section in the case $\mu_2=0$,
a discontinuity is observed at $\mu_1=0$, leading to an infinite conductance.
\modif{%
  At fixed $\mu_1$, the current shows an abrupt jump with $\mu_2$ at small
  dissipation which becomes smoother at larger values of $\gamma$. This
  behavior is understood in the following way: consider for instance the case
  $\mu_1<0$. At zero temperature, the
  quantum state of the left subchain with $\varepsilon_{1,k}=0$ is empty.
  Therefore, a current will flow from right to left only if a fermion
  can undergo a transition from an occupied state of the right subchain.
  In the absence of dissipation, the dynamics is governed only by the
  Hamiltonian. Treating perturbatively the coupling between the two
  subchains, the probability per unit of time of such a transition is
  given at lowest order by the Fermi golden rule. The latter implies
  energy conservation, i.e. a transition can only occur from the state
  of the right subchain for which $\varepsilon_{2,k}=\varepsilon_{1,k}=0$.
  Such a state is
  occupied only if $\mu_2>0$. As a consequence, the current is expected
  to display a jump at $\mu_2=0$. Note that a Dirac distribution
  enforcing the constrain $\varepsilon_{2,k}=0$ is indeed recovered under
  the integral of equation (29):
  ${1\over 4\gamma^2+(\varepsilon_{2,k})^2}\longrightarrow
  {\pi\over 4\gamma^2}\delta(\varepsilon_{2,k})$
  in the limit $\gamma\rightarrow 0$.
  When the system is coupled to thermal baths, equation (29) shows that
  transitions occur from states whose energies are spread around
  $\varepsilon_{2,k}=0$ but with a dispersion $2\gamma$. As a consequence,
  the jump of the current at $\mu_2=0$ is smoothed.
}

\begin{figure}
  \begin{center}
    \psfrag{sigma}[Bc][Bc][1][1]{$G/\gamma g^2$}
    \psfrag{mu}[tc][tc][1][0]{$\mu$}
    \psfrag{gamma=0.01}[Bc][Bc][1][1]{\tiny $\gamma=0.01$}
    \psfrag{gamma=1}[Bc][Bc][1][1]{\tiny $\gamma=1$}
    \psfrag{gamma=10}[Bc][Bc][1][1]{\tiny $\gamma=10$}
    \includegraphics[width=6.37cm]{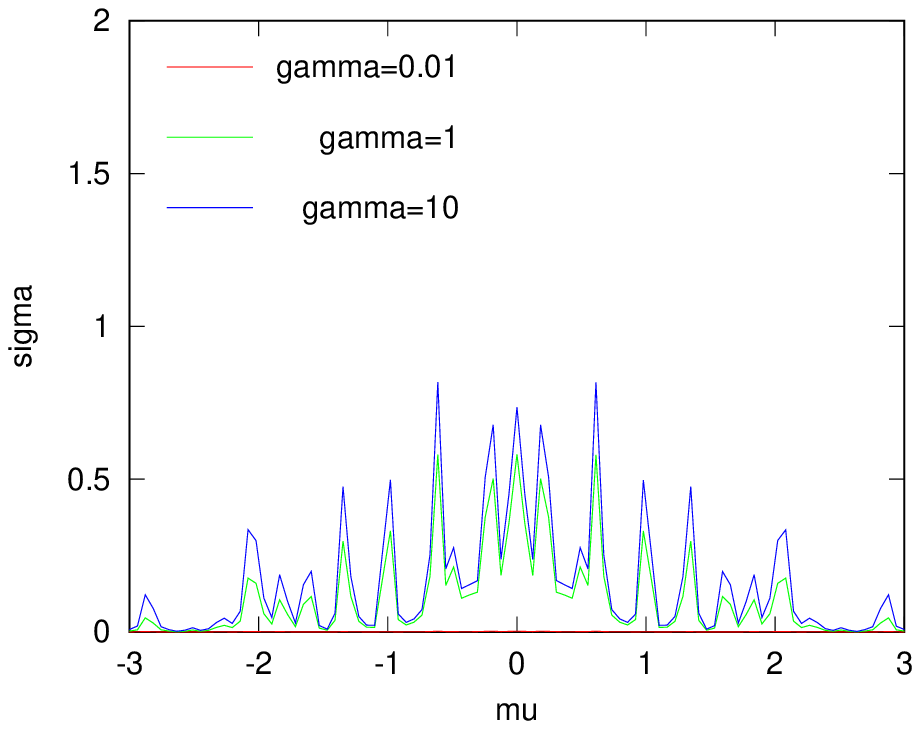}
    \includegraphics[width=6.37cm]{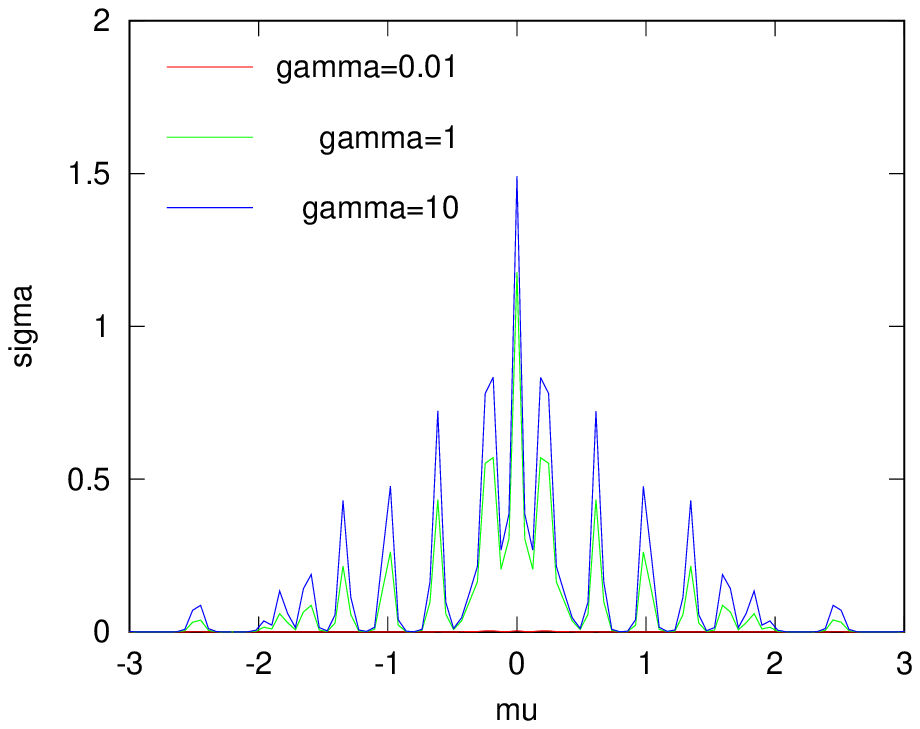}\\
    \includegraphics[width=6.37cm]{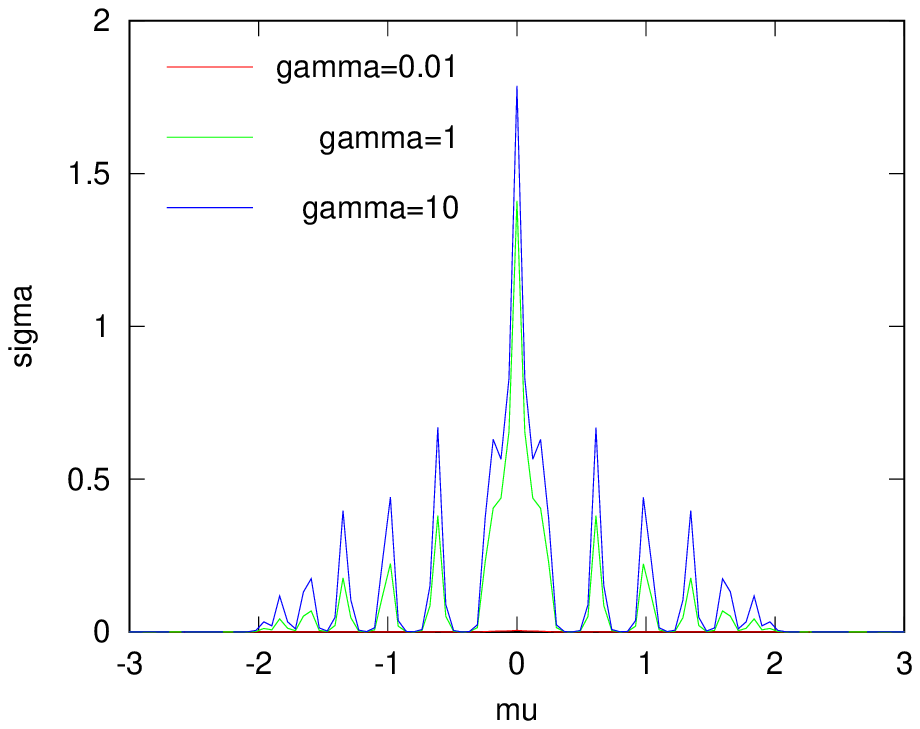}
    \caption{Average conductance $\bar G(\mu)/\gamma g^2$ between a random
      and a pure XX chains of size $L=14$ versus the chemical potential
      $\mu$ when the two chains are in contact with baths at a different
      chemical potentials $\mu+\Delta\mu/2$ and $\mu-\Delta\mu/2$ where
      $\Delta\mu=10^{-4}$.
      The couplings of the random XX model are randomly chosen in $\{1/r;r\}$
      with $r=2$ (top left), $r=4$ (top right) and $r=10$ (bottom).
      The different curves corresponds to different values of the
      dissipation $\gamma$. The temperature is $T=0.02$.}
  \label{fig7}
  \end{center}
\end{figure}

\begin{figure}
  \begin{center}
    \psfrag{sigma}[Bc][Bc][1][1]{$G/\gamma g^2$}
    \psfrag{mu}[tc][tc][1][0]{$\mu$}
    \psfrag{0.01}[Bc][Bc][1][1]{\tiny $\gamma=0.01$}
    \psfrag{g1}[Bc][Bc][1][1]{\tiny $\gamma=1$}
    \psfrag{10}[Bc][Bc][1][1]{\tiny $\gamma=10$}
    \includegraphics[width=6.37cm]{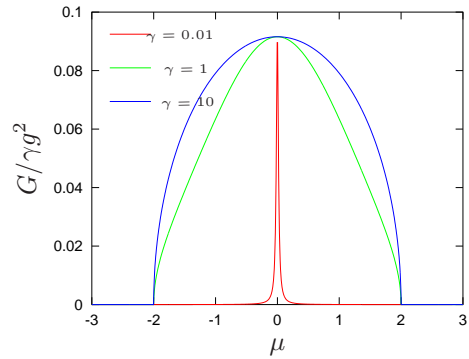}
    \caption{Average conductance $\bar G(\mu)/\gamma g^2$ when the chemical
      potentials of the two reservoirs are $\mu\pm \Delta\mu/2$ versus
      $\mu$ in the infinite-randomness limit. The different curves
      corresponds to different values of the dissipation $\gamma$.
      Note that the Dirac peak at $\mu=0$ has not been plotted.
    }
  \label{fig8}
  \end{center}
\end{figure}

Even though the current does not vanish when $\mu_1\ne 0$, a conductance
can still be defined as the linear response of the system to an imbalance
of the chemical potentials of the two baths. Considering now the setup where
$\mu_1=\mu+\Delta\mu/2$ and $\mu_2=\mu-\Delta\mu/2$, the average conductance
is assumed to be
  \be\bar G(\mu)=\left({d\bar J\over d\Delta\mu}\right)_{\Delta\mu=0}.\ee
Numerically, this quantity was computed by approximating
the derivative with a linear difference:
  \be\bar G(\mu)\simeq {\bar J(\mu+\Delta\mu/2,\mu-\Delta\mu/2)
    -\bar J(\mu,\mu)\over\Delta\mu}\ee
with the small value $\Delta\mu=10^{-4}$. The numerical data are presented
on figures~\ref{fig7}. Due to the discreteness of the spectrum of the
pure subchain, the conductance displays peaks which become sharper and
sharper as the strength of disorder increases. Such a configuration
of peaks is not a specificity of random chains. It is also observed
in pure XX chains (see in particular figure~4 of Ref.~\cite{Pedro}).
However, in the random case, the height of the peak centered at $\mu=0$
diverges as the strength of disorder increases while all other peaks remain
finite. This behavior is understood in the infinite-randomness limit.
Using Eq.~(\ref{Derivee}), the average conductance is found to be
    \be\fl\bar G(\mu)={\gamma g^2\over\pi}\ln{4\over 3}
    {\tan^2k_F\over 1+\gamma^2+\gamma^2\tan^2k_F}{1\over\sin k_F}
    +4\gamma g^2\ln{4\over 3}\left({\sqrt {1+\gamma^{2}}\over\gamma}-1\right)
    \delta(\mu)\ee
where $\mu=-2\cos k_F$. The Dirac distribution at zero chemical potential
discussed in the previous section has been added to the derivative.
Eliminating $k_F$, the expression can be simplified to
    \be\fl\bar G(\mu)={\gamma g^2\over\pi}\ln{4\over 3}
    {4-\mu^2\over 4\gamma^2+\mu^2}{1\over\sqrt{1-\mu^2/4}}
    +4\gamma g^2\ln{4\over 3}\left({\sqrt {1+\gamma^{2}}\over\gamma}-1\right)
    \delta(\mu).\ee
The average conductance is plotted on figure~\ref{fig8}. Though similar
to the one obtained in the pure case when $\gamma=1$, the curve $\bar G(\mu)$
versus $\mu$ is significantly different at low dissipation, i.e. $\gamma\ll 1$.
Since randomness limits the number of channels from one subchain to the
other, a low coupling $\gamma$ with the bath restricts the conductance
to a small window around $\mu=0$.

\section{Conclusions}
A model consisting in two XX chains, one with random couplings
drawn from a binary distribution and the second one homogeneous,
has been considered. The two chains are coupled to different thermal
baths through a non-local Lindblad dissipator and connected by their
edges. Despite the fact that all eigenstates in the random subchain
are localized, a current is induced between the two subchains
in the steady state because of the non-locality of the dissipator.
A divergence of the conductance is observed at zero chemical potential
as the disorder strength increases and is explained in terms of the
density of states of the localized states of the random subchain.

As far as we are aware, such a diverging conductance has not been
observed before in random quantum systems. Our model has indeed
several peculiarities. As mentioned above, the coupling to the baths
is non-local while boundary dissipators are usually considered. 
Then, most studies on Anderson localization are based on tight-binding
models with random local energies rather than random hopping constants.
Moreover, these random energies are usually distributed according to
a Gaussian law or a uniform law. \modif{%
  The distribution of the hopping couplings is a crucial ingredient
  for the divergence of the conductance. The latter is indeed due to
  the fact that, in the infinite-disorder limit, only one eigenstate of
  the fermion Hamiltonian of the random subchain has a
  finite energy while all others have an energy that diverges with the
  strength of disorder. Therefore, a fermion in the right subchain can
  undergo a transition only to this zero-energy state. A variation of
  the chemical potential $\mu_1$ from negative to positive values
  induces an abrupt change of the occupancy
  of this state at zero temperature and therefore a discontinuity of the
  current and a divergence of the conductance. This result is obtained in
  this study with a binary distribution of couplings $r$ and $1/r$ with $r
  \rightarrow +\infty$ but the same result is expected for any random
  Hamiltonian $H=rh$ where $h$ is random matrix with a vanishing determinant.
}

Note that the model considered in this study is not completely
unrealistic. Besides experimental realizations of spin chains, one can
imagine long polymers made of two kinds of monomer with different
electrical conductivities and adsorbed on a metallic surface that
could play the role of a reservoir.

\section*{Acknowledgements}
The authors gratefully thanks Pedro Guimar\~aes for having presented
him his work~\cite{Pedro} during the MECO conference in Lyon.

\end{document}